\documentclass[a4paper,12pt,reqno,superscriptaddress,nofootinbib]{revtex4}
\usepackage{amsmath,amsfonts,amssymb,bm,graphicx,subfigure}
\usepackage[notrig]{physics}
\usepackage{CJKutf8}
\usepackage{orcidlink}
\usepackage{hyperref}
\usepackage[final]{changes}

\UseRawInputEncoding       
\usepackage{comment}

\usepackage{color,soul}

\newcommand{\id}{\textrm{d}}
\let\ve=\varepsilon   
\begin{document}

\begin{CJK*}{UTF8}{gbsn} 

\title{Local Detailed Balance in the Lorenz Model \\\small{Replaces the Butterfly with Frenetic Bursting}} 

\author{ Faezeh Khodabandehlou \orcidlink{0000-0001-8114-6105}, and Christian Maes \orcidlink{0000-0002-0188-697X}\\
{\it Department of Physics and Astronomy, KU Leuven, Belgium}}
\email{christian.maes@kuleuven.be}


\begin{abstract}
The Lorenz system is the canonical low-order model of convective instability, yet
its dissipative and driving terms have never been checked against, nor
constructed from, an explicit thermodynamic bookkeeping. We derive a modification that
satisfies the local-detailed-balance condition for macroscopic
relaxation toward nonequilibrium steady states, thereby identifying the thermodynamic
force,  entropy-production rate and frenesy of the resulting flow. The resulting model produces a transition from a quiescent fixed point to a robust, large-amplitude relaxation oscillation, closely
analogous to recharge-discharge oscillator paradigms used for the
El Ni\~no--Southern Oscillation. The system
alternates between a long, nearly reversible \emph{recharge} phase
and a brief, violently frenetic \emph{discharge} burst, during which
essentially all of the cycle's activity and entropy production is
concentrated.  
\end{abstract}
\maketitle
\end{CJK*}

\section{Introduction}

Lorenz's 1963 truncation of Rayleigh--B\'enard convection~\cite{Lorenz1963}
remains the standard low-order paradigm for the onset of convective
turbulence and, by extension, for discussions of predictability limits in
weather and climate models~\cite{Palmer2000,Ghil2020,GhilChildress1987}. Its three
equations,
\begin{equation}
\dot x=\sigma(y-x),\quad
\dot y=\rho x-y-xz,\quad
\dot z=xy-bz
\label{eq:lorenz}
\end{equation}
are usually presented as a phenomenological Galerkin truncation: $\sigma >0$
is a Prandtl-number ratio, $\rho$ a rescaled Rayleigh number, $b$ a
geometric factor. It is a  drastic, few-degree-of-freedom projection of a genuinely infinite-dimensional system, formally justifiable very close to the onset of convection (small departure from the pure-conduction state). 
 What is rarely asked is whether \eqref{eq:lorenz} is
\emph{thermodynamically consistent} in the precise sense used in modern
nonequilibrium statistical mechanics: does it admit a decomposition into Hamiltonian and dissipative flows driven by an
identifiable thermodynamic force, satisfying local detailed
balance~\cite{MaesLDB2021}, so that entropy production, currents and
dynamical activity (``frenesy''~\cite{MaesFrenesy2020}) are properly identified, and so that linear and nonlinear response theory
around the resulting attractor is licensed by the formalism of nonequilibrium fluctuation--response theory?

This question has immediate relevance for climate science beyond pure
formalism. Any parametrized closure of unresolved convective or
turbulent transport
implicitly assigns a friction and a forcing to slow variables. Only when
consistent with local detailed balance can one legitimately invoke the physically interpretable response
formulas to predict the model's reaction to human interventions or to slow external drifts from unperturbed data. Otherwise such predictions
have no controlled justification.\\

We use the canonical structure of~\cite{MaesNetocny2026} to
show that Lorenz's quadratic nonlinearity is already a legitimate
Hamiltonian flow, and we derive the exact constraint this places on any
admissible dissipative term  (with Lorenz's own $\rho x$ violating it).  We
construct the compliant alternative, including a genuinely non-gradient
piece where an additional symmetry is imposed via Curie's principle that the dissipative (irreversible) part of a system's dynamics cannot break any continuous spatial or structural symmetries belonging to its reversible (Hamiltonian) flow.  We identify the physical ingredients of the new model and how it differs from Lorenz's setup.  We find that the modified dynamics has a steady nonequilibrium attractor in the form of a stable limit cycle. It is reminiscent of the irregular cessation and reversal of the large-scale circulation in turbulent Rayleigh-Bénard convection, theoretically analyzed a decade ago for a dynamics related to the Lorenz equations \cite{Araujo2005}. Here, we are characterizing plume detachment as a genuinely thermodynamic (activity) event.  In fact, in our modified Lorenz model, the local-detailed-balance structure allows to identify entropic and frenetic contributions to the linear response around the nonequilibrium dynamics.\\

\underline{Plan of the paper}:
The next section builds our local-detailed-balance version of the Lorenz dynamics, starting from general (thermodynamically consistent) principles. Section \ref{cha} discusses the change in physical perspective with Section \ref{num} showing the numerical analysis.  We find a transition from a fixed point (conducting state) to a limit cycle, and that is analyzed in Section \ref{aA}.  A single limit cycle shows two different regimes, one quiescent and the other bursting of activity.  We can associate thermodynamic and kinetic interpretations to that by our relying on local detailed balance.  That is worked out in Sections \ref{S:res} and \ref{entfren}, where entropy production and frenesy enter the linear response.  Section \ref{secD} is devoted to a final discussion where we go back to interpreting the conduction-convection transition in our modified Lorenz dynamics.  After the Conclusion, Appendix \ref{ecost} discusses the cost of time-reversal breaking for holding the conducting state. That `needed work' diverges at the (Hopf) transition.

\section{The reversible/dissipative split}

Recent work on the structure of dynamical fluctuations explains how a deterministic equation should look like for macroscopic relaxation to a nonequilibrium steady condition \cite{MaesNetocny2026}.  The main input there is that the fluctuations around that equation satisfy the condition of local detailed balance. The setup starts by selecting a macroscopic variable $M$,  here $M=(x,y,z)$, assumed dynamically autonomous.  The dynamics has a reversible and a dissipative part.  In the present case, it is governed by a Hamiltonian current $J^H(M)$ and a thermodynamic force $F(M)$, with orthogonality condition $J^H(M)\cdot F(M)=0$.  A simple version of  the physical (zero-cost) evolution in ~\cite{MaesNetocny2026}, is then the once closest to {\tt GENERIC} \cite{grm1,grm2,Grmela_2018,GEN}  and takes the form
\begin{equation}
\dot M = J^H(M) + F(M)
\label{eq:generic}
\end{equation}
To build the Hamiltonian flow $J^H$, we take 
\begin{equation}
H(x,y,z)=\tfrac12\big(x^2+y^2+z^2\big),\quad
J=\begin{pmatrix}0&0&0\\0&0&-x\\0&x&0\end{pmatrix}
\label{eq:HJ}
\end{equation}
with $J^{\!\top}=-J$, so that the flow
is generated by an $x$-dependent, angular-momentum-conserving rotation of the
$(y,z)$ pair:
\begin{equation}
J^H(x,y,z) \equiv J\nabla H = (0,\,-xz,\,xy)
\label{eq:JH}
\end{equation}
It reproduces the quadratic terms in \eqref{eq:lorenz} exactly. In that way, the conservation of $x^2+y^2+z^2$ by the Lorenz nonlinearity has become part of the
Hamiltonian structure\footnote{One could say that is bookkeeping vocabulary borrowed from classical mechanics because the mathematical role is the same (generator of a conservative, phase-space-volume-preserving flow).  However, the $(y,z)$ part of $H$ is energy (at least in the Lorenz derivation) originating from an actual energy-conservation law of the underlying convection equations.  The dependence of $H$ in $x$ is for simplicity.}. We interpret $(y,z)$ as the two components of a single rotating order parameter (quadrature components of one complex amplitude) and  we  read $x$ as an order-parameter-type quantity. More on that in Section \ref{cha}.\\

Next, any admissible thermodynamic force $F=(F_x,F_y,F_z)$ must satisfy
$F\cdot J^H=0$:
\begin{equation}
F\cdot J^H = x\big(yF_z - zF_y\big) = 0, \quad \forall\,x
\;\;\Longrightarrow\;\; yF_z = zF_y 
\label{eq:orth}
\end{equation}
Lorenz's convective driving in \eqref{eq:lorenz}, with $F_y$ containing the term $\rho \,x$,  does not satisfy that orthogonality.  That can be remedied and condition \eqref{eq:orth} is satisfied, without being a gradient, by any $F$ of the radial form $(F_y,F_z)=\phi(x,y,z)\,(y,z)$ for some scalar function $\phi$.  The minimal choice is $\phi=\phi(x)$. Therefore, the
general compliant force, leaning most closely to \eqref{eq:lorenz}, is
\begin{equation}
F = \big(\sigma(y- x) - r,\ \ [\phi(x)-1]\,y,\ \ [\phi(x)-1]\,z\big)
\label{eq:Fcompliant}
\end{equation}
for an arbitrary constant $r$ and function $\phi(x)\ge0$.  The factor $\phi(x)-1$ is an $x$-gated, isotropic
self-damping/self-pumping rate acting radially on $(y,z)$. The relaxation equation \eqref{eq:generic} thus 
\begin{align}
\dot x &= \sigma(y-x) - r \notag\\
\dot y &= -xz + [\phi(x)-1]\,y \label{eq:sysz}\\
\dot z &= xy + [\phi(x)-1]\,z \notag
\end{align}
our modified Lorenz-type model.\\
The instantaneous phase-space expansion rate is
\begin{equation}
\nabla\!\cdot\!F = -\sigma + 2[\phi(x)-1] = -(\sigma+2) + 2\phi(x)
\label{eq:div}
\end{equation}
changing sign at $\phi(x^\star)=1+\sigma/2$.  As a specific choice, we take the saturating form
\begin{equation}
\phi(x) = \frac{\kappa x^2}{1+\varepsilon x^2},\qquad \kappa > \varepsilon>0
\label{eq:phi}
\end{equation}
having $\phi(0)=0$, and $\phi(x)\to\kappa/\varepsilon$ as $|x|\to\infty$.
Using \eqref{eq:div}--
\eqref{eq:phi},
\begin{equation}
x^{\star2} = \frac{1+\sigma/2}{\kappa - \varepsilon(1+\sigma/2)}
\label{eq:xstar}
\end{equation}
When $\phi(x) > 1+ \sigma/2$, phase space locally contracts acting on the isotropic $(y,z)$.\\

The equations \eqref{eq:sysz} are not integrable in
closed form, but the $(y,z)$ sector admits an exact, non-perturbative
simplification. Write $y=R\cos\theta$, $z=R\sin\theta$, with
$R=\sqrt{y^2+z^2}\ge0$. Direct substitution gives, without
approximation,
\begin{equation}
\dot R = \big[\phi(x)-1\big]R, \qquad \dot\theta = x,
\label{eq:polar-exact}
\end{equation}
so that \eqref{eq:sysz} is exactly equivalent to
\begin{equation}
\dot x = \sigma(R\cos\theta-x)-r,\qquad
\dot R = [\phi(x)-1]R,\qquad
\dot\theta = x.
\label{eq:polar-system}
\end{equation}
This is not a full solution as $x$ still couples back through
$R\cos\theta=y$, but it isolates two exact facts:\\
1) The rotation rate in $(y,z)$ is exactly $x(t)$;\\
2) For a periodic orbit, $R(T)=R(0)$ with period $T>0$, as $\mathrm{d}(\log R)/\mathrm{d}t=\phi(x)-1$, we get
\begin{equation}
\frac 1{2}\oint_0^T \big[\nabla\!\cdot\!F + \sigma]\,\mathrm \id t = \oint_0^T \big[\phi(x(t))-1\big]\,\mathrm \id t = 0 
\label{eq:quadrature-identity}
\end{equation}
 Since $\nabla\cdot F=-\sigma+2[\phi(x)-1]$, the contribution from the $(y,z)$ sector cancels over one complete cycle, leaving $\oint_0^T\nabla\cdot F\,\mathrm{d}t=-\sigma T$,  or equivalently $\langle\nabla\cdot F\rangle_T=-\sigma$. In other words, the phase-space volume contracts on average at the constant rate $\sigma$ over a complete period.\\  


\section{Interpretation}\label{cha}

Apart from the insertion of $\phi$ and the deleting of $\rho\,x$ for reasons explained above, there are two important changes in \eqref{eq:sysz} from the Lorenz system \eqref{eq:lorenz}:  the appearance of $r\neq 0$, and the symmetric treatment of $y$ and $z$
(where also $b \rightarrow 1-\phi(x)$).\\

 Physically, $r$ is the knob you turn up to inject sustained momentum into the convective mode. For small $r$ the steady condition is conducting with a point attractor.  Above a critical value, as we show below,  the system is forced into a sustained relaxation-oscillation burst cycle.
The route to the limit cycle, a Hopf bifurcation followed by a canard explosion, is exposed in Section \ref{aA}.
This identifies $r$ as the model's actual bifurcation parameter, mechanistically
distinct from Lorenz's $\rho$: $\rho$ destabilizes the origin linearly,  while $r$ instead \emph{displaces} the fixed point away from
the origin and destabilizes it only once the amplitude-gated pumping
$\phi$ catches up with that displacement. \\

The symmetric treatment of $(y,z)$ (with the change $b\rightarrow 1 -\phi(x)$ forced by \eqref{eq:orth}--\eqref{eq:Fcompliant}) is the unique outcome consistent with Curie's symmetry principle for dissipative response.\\ $J^H$ generates an exact
$SO(2)$ rotation of $(y,z)$ at rate $x$.
Curie's principle states that a dissipative force cannot violate a symmetry
the reversible sector respects exactly, unless something in the physical
setup explicitly supplies that symmetry breaking. Here, the friction matrix must commute with the antisymmetric generator $\begin{pmatrix}0&-x\\x&0\end{pmatrix}$, and the only symmetric matrices that do so are multiples of the identity.  Hence, thinking of a single-bath,
isotropic medium, the damping acting on the
$(y,z)$-plane must itself be isotropic, i.e.\ proportional to the identity on
that subspace.  It is worth emphasizing that the orthogonality \eqref{eq:orth} alone would still permit $\phi$ to depend on a direction within the $(y,z)$-plane. Curie's principle, that no dissipative term may resolve a continuous symmetry the reversible sector does not itself break, rules that out, restricting $\phi$ to depend on 
$(x,y^2+z^2)$ at most, and we take the minimal case $\phi=\phi(x)$, making $F_y,F_z$ what they are in \eqref{eq:Fcompliant}.\\
In that same spirit, we  reinterpret $(y,z)$ as the two real components of a single
physical order parameter, rather than Lorenz's two distinct harmonics that
only accidentally appear coupled.  In other words, $(y,z)$ genuinely is one physical
quantity carrying an internal phase.

\subsection{Change in physical picture}

Lorenz's equations \eqref{eq:lorenz} are
obtained from a three-mode Galerkin truncation of two-dimensional Boussinesq
Rayleigh--B\'enard convection between plates held at fixed temperatures
differing by $\Delta T$ \cite{Saltzman1962}. Physically, $x$ is the amplitude of the convective circulation itself
(the intensity of the overturning motion); $y$ is the temperature difference
between the ascending and descending currents (horizontal temperature
contrast, in phase with the velocity mode); and $z$ is the distortion of the
mean vertical temperature profile away from linearity induced by convection.\\  

What is unchanged is that no
alternative truncation of the advective nonlinearity is required to reach \eqref{eq:sysz}, and the buoyancy coupling $\sigma(y-x)$ is likewise retained unchanged
in \eqref{eq:sysz}.  We did not add or redo more microscopic considerations.\\

What is changed, is first of all the treatment of the boundary conditions as a thermodyamic force.  We take a single-bath convective medium,
with no imposed mean temperature gradient; external driving must instead
act directly on the momentum degree of freedom alone, as the compliant
constant force $r$ on $x$ in \eqref{eq:sysz}; e.g.\ a fixed applied body force or mean
pressure gradient, rather than a buoyancy affinity.  Secondly, we require (imposed by  the Curie-principle
requirement motivated above) the $y,z$ to diffuse at the same rate, thinking of the two phase-quadrature components of a \emph{single} Fourier harmonic  (like the real and imaginary parts of the amplitude of a horizontally
\emph{propagating}, rather than standing, convective roll), structurally similar to the device used in deriving the complex Lorenz equations from
baroclinic-instability amplitude equations \cite{GibbonMcGuinness1982}.\\

\subsection{Parity Check}
We can test the above physical picture by a parity check.
Local detailed balance assigns each
variable a definite parity $\varepsilon_i=\pm1$ under time reversal
$w\mapsto \varepsilon w$, $\varepsilon=\mathrm{diag}(\varepsilon_x,\varepsilon_y,\varepsilon_z)$,
fixed by the Onsager--Casimir requirement that the reversible sector be odd
and the dissipative force be even:
\begin{equation}
J^H(\varepsilon w) = -\varepsilon\, J^H(w), \qquad
F(\varepsilon w) = \varepsilon\, F(w)
\end{equation}
A further analysis implies that $\varepsilon_x = +1,\varepsilon_y = +1, \varepsilon_z = -1$:
$x$ and $y$ are time-reversal even, and $z$ is odd.
That is not the parity one would guess from the classical derivation, where
$x$ is a convective-velocity amplitude and would ordinarily be odd. The
constant forcing $r$ is what forces $\varepsilon_x=+1$:
\eqref{eq:Fcompliant} implicitly reinterprets $x$ as an even,
thermodynamic-coordinate-like quantity (an amplitude- or order-parameter-type
variable) rather than a literal velocity. 
The single
even/odd split of $(y,z)$ is exactly what happens to the real and imaginary
parts of one complex mode amplitude under conjugation, announcing the
reading of $(y,z)$ as the two quadrature components of a single rotating
order parameter rather than two independently defined real harmonics.\\

More discussion on the meaning of the variables $x,y,z$ continues in Section \ref{secD}.

\section{Numerical results}\label{num}
We integrate \eqref{eq:sysz} with $\sigma=10$, $\kappa=1$, $\varepsilon=0.05$, $r=15$ (giving $x^\star=2.93$ from \eqref{eq:xstar}), using an adaptive
Runge--Kutta scheme (relative/absolute tolerance $10^{-10}/10^{-12}$).

\begin{figure}[t]
\centering
\includegraphics[width=0.8\linewidth]{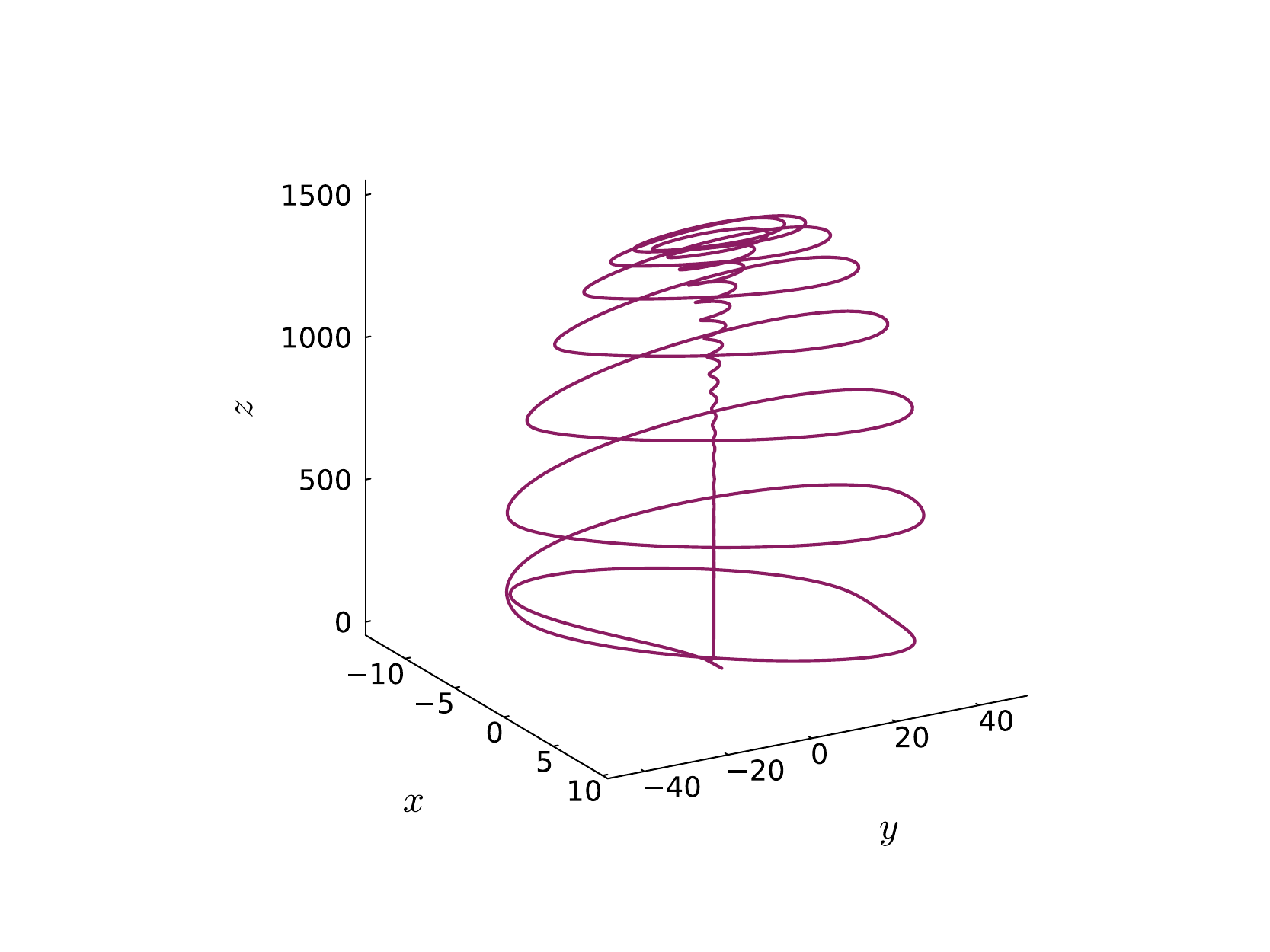}
\caption{Attracting set in $(x,y,z)$ for $\sigma=10$, $\kappa=1$, $\varepsilon=0.05$, $r=15$ in \eqref{eq:sysz} after discarding a transient. The trajectory settles onto a single
closed loop (limit cycle).}
\label{fig:attractor}
\end{figure}
\begin{figure}[t]
\centering
\includegraphics[width=0.6\linewidth]{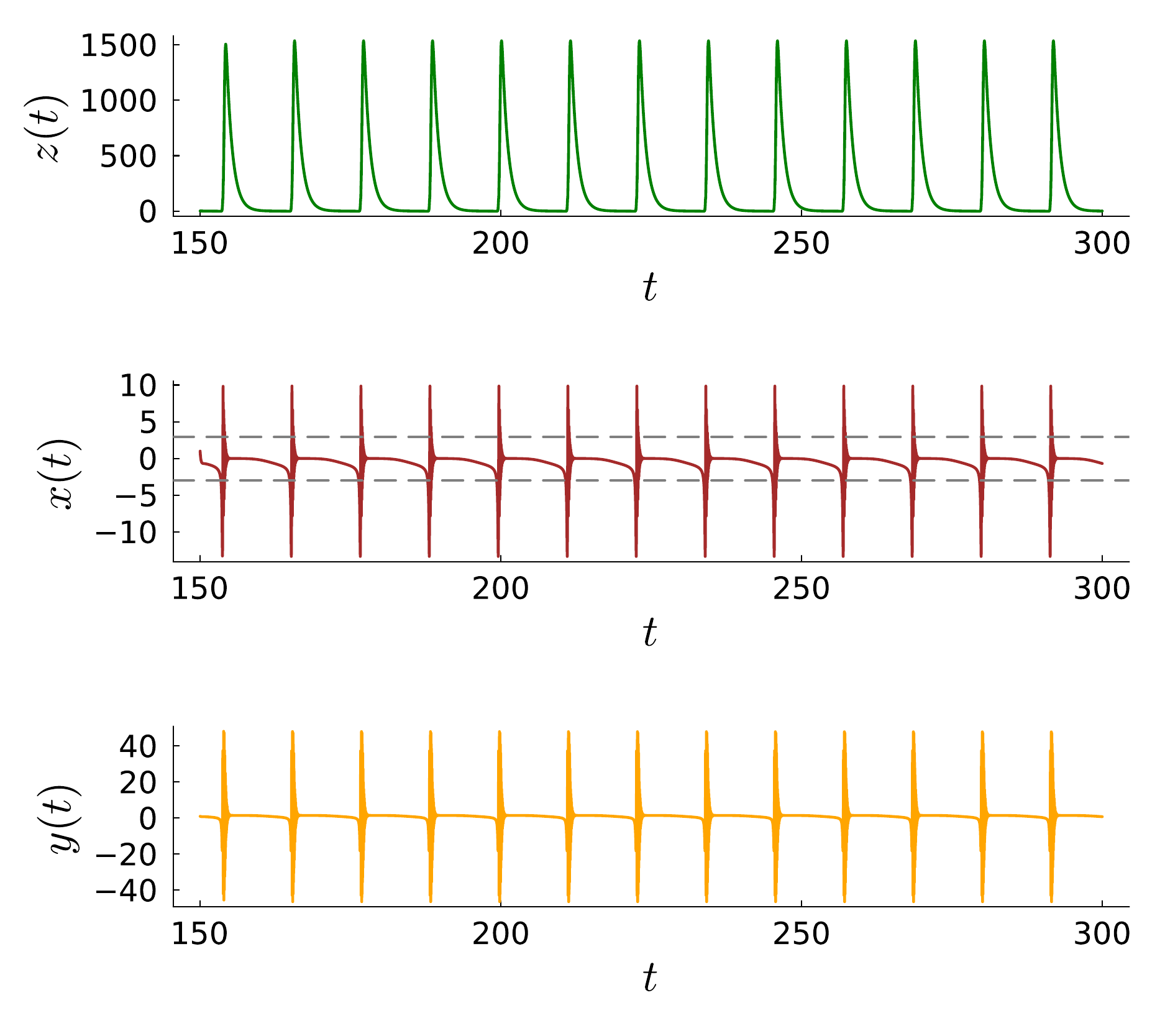}
\caption{Spikes in $z(t)$ recur with  period $T\sim11$, the hallmark of an exactly periodic relaxation oscillation
rather than aperiodic (chaotic) bursting. Note the visibly broken time-reversal symmetry of the waveform itself: each spike rises sharply and decays gradually, a fingerprint of the nonzero entropy production.  Resemblance in shape with recharge-discharge oscillations \cite{Jin1997} is suggestive.}
\label{fig:timeseries}
\end{figure}
Fig.~\ref{fig:attractor} shows the post-transient trajectory: a single closed spatial loop, traversed with amplitude
$x\in[-13.4,9.9]$, $y\in[-46.4,49.1]$, $z\in[-1.1,1537]$, settling in a limit cycle, not the familiar Lorenz
``butterfly.''  In fact,
Fig.~\ref{fig:timeseries} adds evidence of an exactly periodic attractor.\\

During most of each period, $x(t)$ stays confined to a narrow band, roughly between $-3$ and $+3$, in the contraction phase where $x < x^\star\approx 2.93$ (dashed lines): this is precisely the long, nearly reversible recharge phase, with small force, small current, and negligible dissipation.

\begin{figure}[t]
\centering
\includegraphics[width=0.61\linewidth]{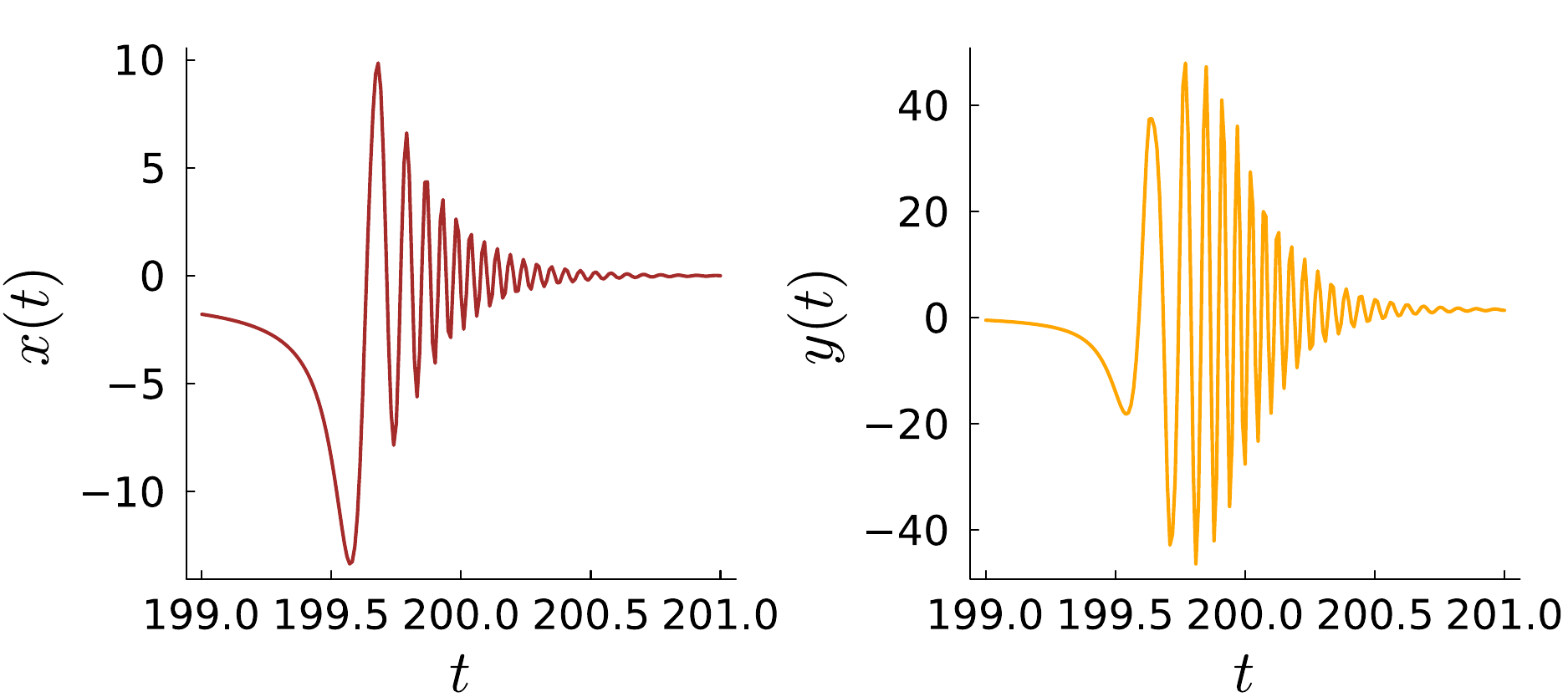}
\caption{$x(t),y(t)$ in a  short time window. }
\label{fig:timeseriesxyzoom}
\end{figure}

In Fig.~\ref{fig:timeseriesxyzoom}, whenever $|x|$ crosses $x^\star$, a sharp discharge burst follows: $x$ shoots up to
$\approx 9.9$ or down to $\approx -13.4$, then relaxes back more slowly.
This rise/decay asymmetry is the broken time-reversal signature.  $y(t)$
follows the same rhythm, with peaks up to $\approx \pm 49$, consistent with
the rotational coupling $\dot y = -xz + [\phi(x)-1]y$: once $x$ crosses
threshold, the $(y,z)$ rotation is briefly and strongly driven.
Each cycle contains both a positive and a negative excursion of $x$ (and
correspondingly of $y$). This is why $z(t)$ in Fig.~2 shows only one type of
spike per period: through $xy$ and the $\phi(x)$ term, $z$ responds to the
\emph{radius} in the $(y,z)$-plane, and is therefore insensitive to the sign
of the excursion in $x$, while $x(t)$ itself visibly alternates between both
signs.\\

The time-averaged phase-space expansion rate along the attractor is $\langle\nabla\!\cdot\!F\rangle_t=-\sigma = -10$. The three Lyapunov exponents measure $\lambda_1=-7.6\times10^{-4}$, $\lambda_2=-3.249$, and $\lambda_3=-6.750$, whose sum, $-9.9996$, agrees with the time-averaged divergence to within $9\times10^{-5}$. The leading exponent is numerically indistinguishable from zero, while the other two are strongly negative, indicating strong transverse contraction toward the observed one-dimensional closed orbit. Together with the closed trajectory and periodic time series in Figs.~\ref{fig:attractor} and \ref{fig:timeseries}, this confirms that the attractor is a stable limit cycle.

\section{Hopf onset and canard-like explosion}
\label{aA}

The numerical results above are all taken at $r=15$. We now trace the emergence of the large-amplitude oscillation as $r$ is increased from zero. The fixed point is unique for every $r$, allowing its stability and the onset of oscillatory behavior to be determined analytically.

\subsection{At the fixed point.} 
Setting $\dot x=\dot y=\dot z=0$ in \eqref{eq:sysz}, the $(y,z)$-equations can be written as
\begin{equation}
\begin{pmatrix}
\phi(x)-1 & -x\\
x & \phi(x)-1
\end{pmatrix}
\begin{pmatrix}
y\\
z
\end{pmatrix}
=0
\end{equation}
 The determinant is $(\phi(x)-1)^2+x^2$, which cannot vanish for the present $\phi(x)$. Hence, $y=z=0$ and there is a unique fixed point
\begin{equation}
M^s=(x_0,y_0,z_0)=(-r/\sigma,0,0)
\label{eq origin}
\end{equation}
Thus, there is no saddle-node multiplicity of fixed points as $r$ is varied.\\

At the fixed point $y_0=z_0=0$, the Jacobian is block triangular
\begin{equation}
J(x_0)=\begin{pmatrix}-\sigma & \sigma & 0\\ 0 & \phi(x_0)-1 & -x_0\\ 0 & x_0 & \phi(x_0)-1\end{pmatrix}
\label{eq:hopf-jacobian}
\end{equation}
so one eigenvalue is $-\sigma$, independently of $r$. The remaining $2\times2$ block has the form $\big[\begin{smallmatrix}a&-b\\b&a\end{smallmatrix}\big]$, with $a=\phi(x_0)-1$ and $b=x_0$, yielding
\begin{equation}
\lambda_{2,3}=\big[\phi(x_0)-1\big]\pm i\,x_0
\label{eqeigrn}
\end{equation}
The imaginary part is therefore set by the rotational coupling of the conservative sector $J^H$: at the fixed point, the linear oscillation frequency is $|x_0|$.

\begin{figure}[t]
\centering
\includegraphics[width=0.65\linewidth]{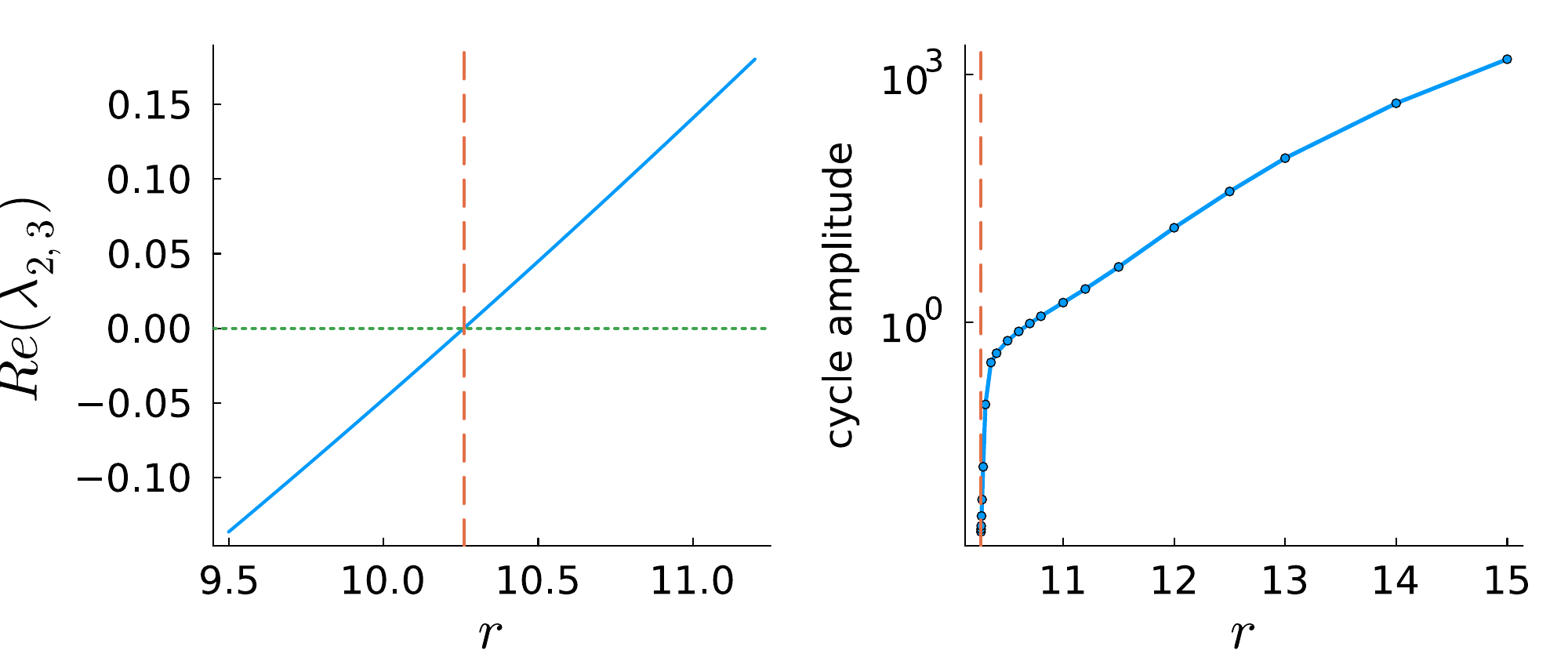}
\caption{(a) Linear stability.  $\mathrm{Re}(\lambda_{2,3})$ from \eqref{eq}, crossing zero at $r_H=10.260$ (dashed). (b) Canard explosion.  Limit-cycle amplitude in $z$ from direct integration started near the fixed point, shown on a logarithmic scale. The amplitude grows continuously from the Hopf onset and subsequently increases very rapidly over a narrow parameter range, indicating a canard-like transition toward large-amplitude relaxation oscillations.}
\label{figlinearstab}
\end{figure}

The complex-conjugate pair crosses the imaginary axis when $\mathrm{Re}(\lambda_{2,3})=0$, namely when $\phi(x_0)=1$:
\begin{equation}
x_0^\star=\frac{1}{\sqrt{\kappa-\varepsilon}},\qquad r_H=\sigma\,x_0^\star=\frac{\sigma}{\sqrt{\kappa-\varepsilon}}
\label{eq}
\end{equation}
where $r_H$ is the linear stability threshold of the fixed point  (but also the point where the 
$(y,z)$-subspace area-contraction rate vanishes, and where the thermodynamic cost of holding the fixed point against noise diverges (see Appendix \ref{ecost}).  For $\sigma=10$, $\kappa=1$, and $\varepsilon=0.05$, we get $r_H=10.260$. The corresponding eigenvalues have nonzero imaginary part, $\mathrm{Im}(\lambda_{2,3})=\pm1.026$, and the real part changes sign as $r$ passes through $r_H$, confirmed in Fig.~\ref{figlinearstab}(a).

\subsection{Canard-like transition.} 
\begin{figure}[t]
\centering
\includegraphics[width=1.01\linewidth]{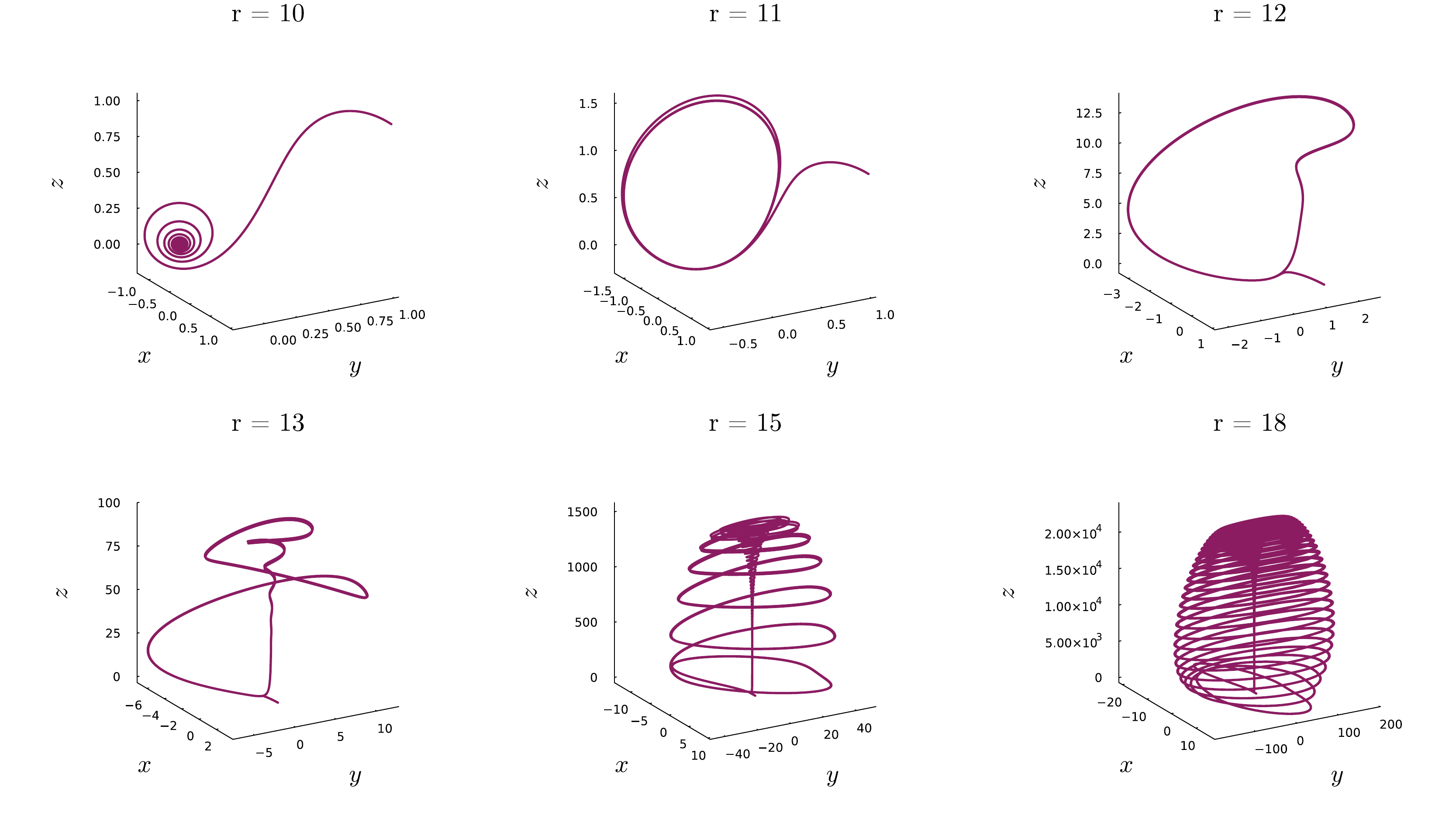}
\caption{\small{The limit cycle in $(x,y,z)$ for six values of $r>r_H=10.260$.
At $r=11$, just above threshold, it is a small, near-elliptical
Hopf-born loop. The canard-explosion is visible from the
amplitude growth: axis scales differ by four orders of magnitude across panels
($z_{\max}\approx2$ at $r=11$ to $z_{\max}\approx9\times10^4$ at
$r=20$).}}
\label{fig:limitcycles-vs-r}
\end{figure}

Table~\ref{tabr} and Fig.~\ref{figlinearstab}(b) show the rapid growth of the limit-cycle amplitude above $r_H$. Very close to the Hopf point, the emerging oscillations are small, while the amplitude increases strongly as $r$ is increased. The transition from ``quiescent fixed point'' to ``explosive 
 relaxation bursts'' happens over a very narrow range of $r$, even though the underlying bifurcation itself is soft; see Fig.~\ref{fig:limitcycles-vs-r}.  We will interpret them below as frenetic bursts.\\
 The initial growth is consistent with the small-amplitude scaling expected near a supercritical Hopf bifurcation, while the subsequent rapid increase suggests a canard-like transition toward relaxation-oscillation behavior. The fast--slow structure of the equations provides a natural mechanism for such behavior: $\sigma=10$ makes $x$ relax rapidly relative to the $(y,z)$ sector.
\begin{table}[t]
\centering
\begin{tabular}{c c c}
\hline
$r$ & $r-r_H$ & amplitude in $z$ \\
\hline
$10.260$ & $0.0002$ & $0.003$ \\
$10.30$ & $0.04$ & $0.05$ \\
$10.60$ & $0.34$ & $0.78$ \\
$11.0$ & $0.74$ & $1.7$ \\
$12.0$ & $1.74$ & $14$ \\
$13.0$ & $2.74$ & $98$ \\
$15.0$ & $4.74$ & $1538$ \\
\hline
\end{tabular}
\caption{Limit-cycle amplitude in $z$ versus distance above the Hopf threshold $r_H=10.260$, obtained by direct integration starting near the fixed point \eqref{eq}.}
\label{tabr}
\end{table}

As the cycle grows, it eventually explores the larger phase-space-expansion threshold $x^\star=2.93$ of \eqref{eq:xstar} for $\sigma = 10, \kappa = 1, \ve = 0.05$, which is distinct from the Hopf threshold $x_0^\star=1.026$. The radial dynamics can produce large excursions. At larger forcing, including $r=15$, the attractor has developed into the fast-rise/slow-decay relaxation oscillation shown in Fig.~\ref{fig:limitcycles-vs-r}.  There we see 3D phase-space trajectories for $t\in[0,300]$ at different values of $r$ at $\ve=0.05, \kappa=1$. For $r<r_H$, the dynamics converge to a point attractor at the origin. As $r$ increases through $r_H$, a limit cycle is created, and the trajectory approaches a closed periodic orbit that is repeated indefinitely.

\section{Linear response formula}\label{S:res}
The canonical Lagrangian (Eq.~III.8 of \cite{MaesNetocny2026}) corresponding to the modified Lorenz model \eqref{eq:sysz} is the Onsager--Machlup rate
function,
\begin{equation}
\mathcal L(M,\dot M) = \frac14\big(\dot M-J^H(M)-F(M)\big)^2
\label{eq:OM-lagrangian}
\end{equation}
Perturbing $F\to F+\delta F$ gives, to first order,
\begin{equation}
\delta\mathcal L(M,\dot M) = \frac12\,\delta F(M)\cdot\big[F(M)+J^H(M)-\dot M\big]
\label{eq:dL}
\end{equation}
The Lagrangian \eqref{eq:OM-lagrangian} governs the path-probability
weight of the diffusion
\begin{equation}
\dot M(t) = \dot M_{\rm typ}(t) + \sqrt2\,\xi_t
\label{eq:finite-N-sde}
\end{equation}
and $\dot M(s)-\dot M_{\rm typ}(s)=\sqrt2\,\xi(s)$ is a genuinely
fluctuating quantity (white noise).  The time-correlation
$\langle O(t)\,[\dot M(s)-\dot M_{\rm typ}(s)]\rangle_0$
depends on the path over the whole interval $[s,t]$  and measures the sensitivity of the (later)
observable $O(t)$ to a kick received at time $s$.\\
Operationally, this sensitivity is simplest to access
directly~\cite{BaiesiMaesWynants2009}. Simulate the finite-noise
process \eqref{eq:finite-N-sde}, record the actual increments
$\Delta M(s)=M(s+\Delta s)-M(s)$, which do not equal
$\dot M_{\rm typ}(s)\Delta s$ exactly precisely because the noise is
retained, and correlate them with $O(t)$ empirically over many
noisy realizations.\\
In that way we find the linear response 
\begin{equation}
\delta\langle O(t)\rangle_t = \frac12\int_0^t\!\mathrm ds\,
\Big\langle O(t)\,\delta F\cdot\big[\dot M(s)-\dot M_{\rm typ}(s)\big]\Big\rangle_0
\label{eq:response}
\end{equation}
where $\dot M_{\rm typ}\equiv F(M)+J^H(M)$ is the right-hand side of
the zero-cost flow~\eqref{eq:generic}, the local, deterministic drift,
and the correlation is evaluated on the process~\eqref{eq:finite-N-sde}.
Equation~\eqref{eq:response} is the correct nonequilibrium
Green--Kubo-type formula licensed by local detailed
balance~\cite{BaiesiMaesWynants2009}: response is governed by the
correlation of $O$ with the deviation of the realized current from
the typical drift, not by the current itself.\\
Concretely, for a perturbation $\delta F=(\delta r,0,0)$,
\eqref{eq:response} reduces to the time-correlation
\begin{equation}
\delta\langle O(t)\rangle_t = \frac{\delta r}{2}\int_0^t\!\mathrm ds\,
\Big\langle O(t)\,\Big[\dot x(s)-\sigma[y(s)-x(s)]+r\Big]\Big\rangle_0
\label{eq:response-r}
\end{equation}
fully explicit and computable from the already-known limit-cycle
trajectory, without rerunning the perturbed dynamics.\\

Eqs~\eqref{eq:response} and \eqref{eq:response-r} hold
unchanged for any $r$.  However, there is a genuinely qualitative difference. Every direction transverse to a stable fixed point
is contracting, so $\delta\langle O(t)\rangle_t\to0$ as $t\to\infty$
for generic $O$: any perturbation eventually relaxes away. On the other hand, the limit
cycle has the tangent direction $v(t)=\dot M_{\rm typ}(t)$, associated
with the cycle's own time-translation invariance, with no restoring force at all. A perturbation
$\delta F$ with a component along this direction shifts the phase of
the oscillation rather than decaying: the corresponding part of
$\delta\langle O(t)\rangle_t$, for $O$ sensitive to phase, need not
relax to a finite limit but can persist (subject to the
phase-diffusion smearing that noise introduces over long times).\\

Exactly at the Hopf threshold, the linear response has a sharp
signature. The static (DC) susceptibility stays
finite but the resonant response diverges.
Driving the system at the very
frequency it is about to start oscillating at spontaneously produces
an unbounded response. Correspondingly, the step response to a
constant $\delta F$ stops relaxing to its (finite) steady value as
$r\to r_H$ and the relaxation time diverges (critical slowing
down). At $r=r_H$ itself the response oscillates forever.

\section{Entropy production and frenesy on the limit cycle}\label{entfren}

The model describes a transition from a conducting state to a convective condition.  There is of course entropy production in the conducting state as well, as there is a heat current.  However, the present model (as the original Lorenz model) on the macroscopic scale does not consider that variable; rather, the coarse-graining involves the convective current which is absent in the conducting state and does not contribute to the considered entropy production; see however Appendix \ref{ecost}.\\

When crossing the transition, we have a nontrivial attractor where the state returns to itself exactly
after one period $T$. The entropy exported to the environment,
$\oint_0^T j(s)\cdot F(z_s)\,\mathrm ds$, has become strictly positive:
evaluated numerically along the attractor of Fig.~\ref{fig:attractor},
\begin{equation}
\oint_0^T j\cdot F\,\mathrm ds \;\approx\; 8.8\times10^{6} \;>\;0
\end{equation}
consistently with the Second Law and with monotonicity of the
nonequilibrium entropy, Eq.~(II.9) of \cite{MaesNetocny2026}.  We have a genuine current carrying nonequilibrium condition, continuously dissipating heat supplied by the
constant thermodynamic force $r$ doing work against the damping channels,
at a rate that exactly balances over each cycle so the orbit closes.\\

This entropy production is not only computable but also reflected in the time series: the waveform in Fig.~2 rises sharply and decays gradually. The $5\%$-to-$95\%$ rise time of a spike is $0.44$, whereas the $95\%$-to-$5\%$ decay time is $2.72$, giving a factor of $6.2$ asymmetry. This pronounced temporal asymmetry is consistent with the broken time-reversal symmetry already quantified thermodynamically by $\oint j\cdot F,\mathrm ds>0$. The fast rise is associated with threshold crossing, where $\phi(x)-1>0$ and the $(y,z)$ amplitudes are locally amplified. By contrast, once $x$ falls below $x^\star$, $\phi(x)-1<0$ and the dynamics becomes relaxation-dominated; as $x$ decreases further, $|\phi(x)-1|\to1$. The recharge and discharge phases therefore correspond to distinct dynamical regimes with different characteristic time scales, producing the strongly asymmetric waveform.\\

Frenesy, $\propto|F(z)|^2$ here, is wildly
inhomogeneous along the cycle: evaluated along the same trajectory it
ranges over eight orders of magnitude, from $|F|^2\approx0.6$ during the
quiescent phase (small $x,y,z$, negligible activity) to
$|F|^2\approx7.3\times10^7$ at the peak of the burst ($|x|>x^\star$,
$z\sim900$) — a ratio of order $10^8$. Physically, the attractor
alternates between a long, nearly frenesy-free \emph{recharge} phase,
during which the system sits close to the origin and the dynamics is
almost reversible (tiny force, tiny current, negligible dissipation),
and a brief, violently frenetic \emph{discharge} burst, during which
essentially all of the cycle's activity and entropy production is
concentrated. This is the thermodynamic signature underlying the
recharge-discharge phenomenology already invoked for
ENSO~\cite{Jin1997}: the relaxation oscillator $(x,y,z)$ is a system that spends
almost all of its time in an inactive, low-dissipation state and
periodically pays its entire thermodynamic cost in one short, intense
event.

\section{Discussion}\label{secD}
Lorenz's motivation in 1963 was probably not to build an accurate convection model. Rather, it was to show that a simple, fully deterministic system of ODEs could nonetheless exhibit sensitive dependence on initial conditions, a phenomenon that became known as the ``butterfly effect.'' The three-mode truncation became the standard minimal example used across the whole field of nonlinear dynamics to exhibit a strange attractor and deterministic chaos, taught and cited independently of whether it says anything quantitatively correct
about actual fluid convection. which, as discussed in the Introduction, it does not~\cite{Curry1978,CrossHohenberg1993}.\\

Sufficiently close to the onset of convection, a systematic weakly-nonlinear expansion (not Lorenz's {\it ad hoc} truncation, but a reduction asymptotically consistent in the distance from the critical Rayleigh number) correctly recovers the pitchfork bifurcation from conduction to steady convection rolls, with the real Ginzburg--Landau equation governing the slowly varying envelope of the emerging rolls~\cite{NewellWhitehead1969,Segel1969}. Retaining genuine spatial structure rather than the envelope alone, the Swift--Hohenberg
equation reproduces how a real, bounded fluid layer selects a specific roll wavelength at onset~\cite{SwiftHohenberg1977,CrossHohenberg1993}; more complete multi-mode Galerkin treatments confirm this picture and push the onset of chaotic behavior to far higher, more physically realistic temperature gradients than Lorenz's three-mode truncation
suggests~\cite{Curry1978,CurryHerringLoncaricOrszag1984,GhilChildress1987}.\\
In that context, our findings above are not detailing a more motivated derivation of a (modified) Lorenz dynamics.  Our starting point remains the Lorenz system and we have applied general principles to put it into the structure \cite{MaesNetocny2026} of macroscopic equations in a fluctuating world satisfying local detailed balance. For that reason the interpretation of the variables and the parameters in \eqref{eq:lorenz} are relevant for the new model itself.\\

A natural reading is to say that $x$ is no longer the convective-roll amplitude itself (which would have odd parity), but rather a mean energy at some vertical location.  The $\phi(x)$ is then seen as an energy/temperature-dependent viscosity or diffusivity.\\
 The base state $(x^*,y^*,z^*)=(-r/\sigma,0,0)$ is a purely linear, conduction-like profile (analogue of the classical no-convection state), but now it exists for every value of $r$.  Its stability is governed not by $r$ directly but by whether the induced current $x^*=-r/\sigma$ has crossed the threshold $|x^*|=1/\sqrt{\kappa-\varepsilon}$. Crossing that threshold is exactly the Hopf point $r_c=\sigma/\sqrt{\kappa-\varepsilon}$, and can be seen as the model's analogue of the critical Rayleigh number. However, here the onset is a  Hopf bifurcation straight into oscillatory/bursting behavior. That is structurally closer to the onset of oscillatory convection seen in more realistic settings \cite{Chandrasekhar1961,Veronis1965,HurleJakeman1971,KolodnerBensimonSurko1988,Radko2013,Krishnamurti1973,Krishnamurti1981}.
The subsequent rapid amplitude growth  then reads as the onset of intermittent, bursty convection with long, quiet, near-conductive intervals punctuated by violent convective overturning events such as discussed in \cite{Araujo2005}.  That is qualitatively reminiscent of real intermittent plume-bursting convective regimes  rather than the smooth, steady rolls of the textbook Rayleigh–Bénard picture.\\

While
thermodynamic consistency (local detailed balance combined with the Curie principle) appears to destroy chaos, the periodic oscillation obtained instead is not an artifact:
ENSO recharge-discharge oscillators~\cite{Jin1997,Jiang1995} are, in essence,
low-order relaxation oscillators of exactly this
build-threshold-relax-repeat character. Ours comes with the full
nonequilibrium bookkeeping (force, current, entropy production,
frenesy) attached from the outset, licensing the response formula
\eqref{eq:response} for its sensitivity to slow external drift.  We believe that such procedure is therefore of
direct relevance to detection and attribution of forced change in
oscillatory climate modes.

\section{Conclusion}
We have not derived a model for the transition from conduction to convection starting from higher-order equations. Rather, we took the Lorenz equations directly and examined a modification consistent with local detailed balance and with the symmetry between the reversible and dissipative flow.\\
Within the family \eqref{eq:phi}, we find that thermodynamic consistency converts Lorenz's chaotic convective instability into a robust, self-sustained periodic relaxation oscillation. Beyond its bearing on Lorenz's own equations, the orthogonality $F\cdot J^H=0$ used throughout applies unchanged to any low-order closure, so the resulting oscillator is of independent interest for its recharge-discharge phenomenology.\\
The resulting dynamics exhibits a Hopf transition from a conducting state to a limit cycle, with bursts of activity interspersed between largely reversible, quiescent intervals. Local detailed balance further allows us to give this behavior both entropic and frenetic interpretations.

\vspace{1cm}
\noindent {\bf Acknowledgment}: FK is supported by the Research Foundation–Flanders (FWO) through postdoctoral fellowship 1232926N and acknowledges  useful discussions during a one-month visit to the Isaac Newton Institute (Cambridge, UK); special thanks to Valerio Lucarini and Niccolò Zagli.

\appendix

\section{Cost of time-reversal breaking \\in the normal approximation around the fixed point}
\label{ecost}
 
For $\dot z=Az+\sqrt{2D}\,\xi$ with stationary Gaussian density
$\rho^\text{stat}\propto e^{-\frac12z^TC^{-1}z}$, the physically correct
reversed path is $w(t)=\varepsilon z(T-t)$ over a period $[0,T]$. One can use the idea and calculations of \cite{2000} to associate an entropy production to that process. We summarize the findings.\\

Applying this to the $2\times2$ $(y,z)$ block alone, with the
correct sub-parity $\varepsilon_2=\mathrm{diag}(1,-1)$ ($y$ even,
$z$ odd) yields zero entropy production: the rotational
current in $(y,z)$ is an exactly
reversible, equilibrium-like oscillation once $y,z$ are treated as
the position- and momentum-like quadrature components.  That is nothing else than a correctly-paired
position/momentum oscillator satisfying detailed balance with respect
to its own stationary Gaussian regardless of its damping rate.\\
 In fact, the ``entropy production'' (as measure of time-reversal breaking) comes entirely from the $x$-coupling.  We find it to be
\begin{equation}
 \frac{\sigma^2\big(a^2-a\sigma+x_0^2\big)}
{-a\big[(a-\sigma)^2+x_0^2\big]}, \qquad a=\phi(x_0)-1,\; x_0 =  -r/\sigma
\label{eq:Pi-eps-full}
\end{equation}
manifestly positive for $a<0$ (stability). Since the $(y,z)$ sector
alone contributes nothing, \eqref{eq:Pi-eps-full} is entirely
generated by the $x$--$(y,z)$ coupling $\sigma(y-x)-r$: it is the
driving $r$, transmitted through $\sigma$, that costs
entropy, not the rotation it drives.\\
Note the divergence in \eqref{eq:Pi-eps-full} with $1/|a|$: the thermodynamic cost of holding the point attractor
diverges together with its linear stability.

\bibliographystyle{apsrev4-2}
\bibliography{Lorenz}

@article{Lorenz1963,
  author  = {Lorenz, E.  N.},
  title   = {Deterministic Nonperiodic Flow},
  journal = {Journal of the Atmospheric Sciences},
  volume  = {20},
  number  = {2},
  pages   = {130--141},
  year    = {1963},
  doi     = {10.1175/1520-0469(1963)020<0130:DNF>2.0.CO;2}
}

@article{Araujo2005,
  author  = {Araujo, F. F. and Grossmann, S. and Lohse, D.},
  title   = {Wind Reversals in Turbulent {R}ayleigh-{B}{\'e}nard Convection},
  journal = {Phys. Rev. Lett.},
  volume  = {95},
  number  = {8},
  pages   = {084502},
  year    = {2005},
  doi     = {10.1103/PhysRevLett.95.084502}
}

@article{Palmer2000,
  author  = {Palmer, T. N.},
  title   = {Predicting uncertainty in forecasts of weather and climate},
  journal = {Reports on Progress in Physics},
  volume  = {63},
  number  = {2},
  pages   = {71--116},
  year    = {2000},
  doi     = {10.1088/0034-4885/63/2/201}
}

@article{NewellWhitehead1969,
  author  = {Newell, Alan C. and Whitehead, J. A.},
  title   = {Finite bandwidth, finite amplitude convection},
  journal = {Journal of Fluid Mechanics},
  volume  = {38},
  number  = {2},
  pages   = {279--303},
  year    = {1969},
  doi     = {10.1017/S0022112069000176}
}

@article{Segel1969,
  author  = {Segel, Lee A.},
  title   = {Distant side-walls cause slow amplitude modulation of cellular convection},
  journal = {Journal of Fluid Mechanics},
  volume  = {38},
  number  = {1},
  pages   = {203--224},
  year    = {1969},
  doi     = {10.1017/S0022112069000127}
}

@article{Grmela_2018,
doi = {10.1088/2399-6528/aab642},
url = {https://dx.doi.org/10.1088/2399-6528/aab642},
year = {2018},
publisher = {IOP Publishing},
volume = {2},
number = {3},
pages = {032001},
author = {Grmela, M.},
title = {{GENERIC guide to the multiscale dynamics and thermodynamics}},
journal = {Journal of Physics Communications}
}

@article{grm1,
author  = {M. Grmela and H. C. {\"O}ttinger},
title   = {Dynamics and thermodynamics of complex fluids. I. Development of a general formalism},
journal = {Physical Review E},
volume  = {56},
pages   = {6620},
year    = {1997}
}

@article{grm2,
author  = {M. Grmela and H. C. {\"O}ttinger},
title   = {Dynamics and thermodynamics of complex fluids. II. Illustrations of a general formalism},
journal = {Physical Review E},
volume  = {56},
pages   = {6633},
year    = {1997}
}

@book{GEN,
author    = {H. C. {\"O}ttinger},
title     = {Beyond Equilibrium Thermodynamics},
publisher = {Wiley-Interscience},
address   = {New York},
year      = {2005}
}

@article{SwiftHohenberg1977,
  author  = {Swift, J. and Hohenberg, P. C.},
  title   = {Hydrodynamic fluctuations at the convective instability},
  journal = {Physical Review A},
  volume  = {15},
  number  = {1},
  pages   = {319--328},
  year    = {1977},
  doi     = {10.1103/PhysRevA.15.319}
}

@article{CrossHohenberg1993,
  author  = {Cross, M. C. and Hohenberg, P. C.},
  title   = {Pattern formation outside of equilibrium},
  journal = {Reviews of Modern Physics},
  volume  = {65},
  number  = {3},
  pages   = {851--1112},
  year    = {1993},
  doi     = {10.1103/RevModPhys.65.851}
}

@article{Curry1978,
  author  = {Curry, J. H.},
  title   = {A generalized {L}orenz system},
  journal = {Communications in Mathematical Physics},
  volume  = {60},
  number  = {3},
  pages   = {193--204},
  year    = {1978},
  doi     = {10.1007/BF01612888}
}

@article{CurryHerringLoncaricOrszag1984,
  author  = {Curry, J. H. and Herring, J. R. and Loncaric, J.  and Orszag, S.  A.},
  title   = {Order and disorder in two- and three-dimensional {B}\'enard convection},
  journal = {Journal of Fluid Mechanics},
  volume  = {147},
  pages   = {1--38},
  year    = {1984},
  doi     = {10.1017/S0022112084001968}
}

@book{Chandrasekhar1961,
  author    = {Chandrasekhar, S.},
  title     = {Hydrodynamic and Hydromagnetic Stability},
  publisher = {Clarendon Press},
  address   = {Oxford},
  year      = {1961}
}

@article{Veronis1965,
  author  = {Veronis, G. },
  title   = {On finite amplitude instability in thermohaline convection},
  journal = {Journal of Marine Research},
  volume  = {23},
  number  = {1},
  pages   = {1--17},
  year    = {1965}
}

@article{HurleJakeman1971,
  author  = {Hurle, D. T. J. and Jakeman, E.},
  title   = {Soret-driven thermosolutal convection},
  journal = {Journal of Fluid Mechanics},
  volume  = {47},
  number  = {4},
  pages   = {667--687},
  year    = {1971},
  doi     = {10.1017/S0022112071001284}
}

@article{KolodnerBensimonSurko1988,
  author  = {Kolodner, P. and Bensimon, D. and Surko, C. M.},
  title   = {Travelling-wave convection in an annulus},
  journal = {Physical Review Letters},
  volume  = {60},
  number  = {17},
  pages   = {1723--1726},
  year    = {1988},
  doi     = {10.1103/PhysRevLett.60.1723}
}

@book{Radko2013,
  author    = {Radko, T. },
  title     = {Double-Diffusive Convection},
  publisher = {Cambridge University Press},
  address   = {Cambridge},
  year      = {2013},
  doi       = {10.1017/CBO9781139034173}
}

@article{Ghil2020,
  author  = {Ghil, M.  and Lucarini, V. },
  title   = {The physics of climate variability and climate change},
  journal = {Reviews of Modern Physics},
  volume  = {92},
  pages   = {035002},
  year    = {2020},
  doi     = {10.1103/RevModPhys.92.035002}
}

@book{MaesLDB2021,
  author    = {Maes, C. },
  title     = {Local Detailed Balance},
  series    = {SciPost Physics Lecture Notes},
  volume    = {32},
  year      = {2021},
  publisher = {SciPost},
  doi       = {10.21468/SciPostPhysLectNotes.32}
}

@article{MaesFrenesy2020,
  author  = {Maes, C. },
  title   = {Frenesy: Time-symmetric dynamical activity in nonequilibrium systems},
  journal = {Physics Reports},
  volume  = {850},
  pages   = {1--33},
  year    = {2020},
  doi     = {10.1016/j.physrep.2020.01.002}
}

@article{2000,
author = {C. Maes and F. Redig and A. Van Moffaert},
title = {On the definition of entropy production via examples},
journal = {J. Math. Phys.},
volume = {41},
pages = {1528--1554},
year = {2000}
}

@article{MaesNetocny2026,
  author  = {Maes, Christian and Neto{\v{c}}n{\'y}, Karel},
  title   = {Relaxation to nonequilibrium},
  journal = {Journal of Non-Equilibrium Thermodynamics},
  year    = {2026},
  doi     = {10.1515/jnet-2026-0037}
}

@article{BaiesiMaesWynants2009,
  author  = {Baiesi, M.  and Maes, C.  and Wynants, B. },
  title   = {Fluctuations and Response of Nonequilibrium States},
  journal = {Physical Review Letters},
  volume  = {103},
  pages   = {010602},
  year    = {2009},
  doi     = {10.1103/PhysRevLett.103.010602}
}

@article{Jin1997,
  author  = {Jin, Fei-Fei},
  title   = {An Equilibrium Theory for Atmospheric Teleconnections},
  journal = {Journal of the Atmospheric Sciences},
  volume  = {54},
  number  = {6},
  pages   = {811--829},
  year    = {1997},
  doi     = {10.1175/1520-0469(1997)054<0811:AETFAT>2.0.CO;2}
}

@article{Saltzman1962,
  author  = {Saltzman, Barry},
  title   = {Finite Amplitude Free Convection as an Initial Value Problem---I},
  journal = {Journal of the Atmospheric Sciences},
  volume  = {19},
  number  = {4},
  pages   = {329--341},
  year    = {1962},
  doi     = {10.1175/1520-0469(1962)019<0329:FAFCAA>2.0.CO;2}
}

@article{GibbonMcGuinness1982,
  author  = {Gibbon, J. D. and McGuinness, M. J.},
  title   = {The complex Lorenz equations},
  journal = {Physica D: Nonlinear Phenomena},
  volume  = {5},
  number  = {1},
  pages   = {108--122},
  year    = {1982},
  doi     = {10.1016/0167-2789(82)90010-0}
}

@article{Krishnamurti1981,
  author  = {Krishnamurti, R.  and Howard, Louis N.},
  title   = {Large-scale flow generation in turbulent convection},
  journal = {Proceedings of the National Academy of Sciences of the United States of America},
  year    = {1981},
  volume  = {78},
  number  = {4},
  pages   = {1981--1985},
  doi     = {10.1073/pnas.78.4.1981}
}

@article{Krishnamurti1973,
  author  = {Krishnamurti, Ruby},
  title   = {Some further studies on the transition to turbulent convection},
  journal = {Journal of Fluid Mechanics},
  year    = {1973},
  volume  = {60},
  number  = {2},
  pages   = {285--303},
  doi     = {10.1017/S0022112073000170}
}

@book{GhilChildress1987,
  author    = {Ghil, M. and Childress, S. },
  title     = {Topics in Geophysical Fluid Dynamics: Atmospheric Dynamics, Dynamo Theory and Climate Dynamics},
  series    = {Applied Mathematical Sciences},
  volume    = {60},
  publisher = {Springer},
  address   = {New York},
  year      = {1987},
  pages     = {xv + 485},
  isbn      = {978-0-387-96475-1},
  doi       = {10.1007/978-1-4612-1052-8},
  note      = {Reissued as an eBook by Springer, 2012, ISBN 978-1-4612-1052-8}
}

@article{Jiang1995,
  author  = {Jiang, S.  and Jin, F. F.  and Ghil, M. },
  title   = {Multiple equilibria, periodic, and aperiodic solutions in a wind-driven, double-gyre, shallow-water model},
  journal = {Journal of Physical Oceanography},
  year    = {1995},
  volume  = {25},
  number  = {5},
  pages   = {764--786},
  doi     = {10.1175/1520-0485(1995)025<0764:MEPAAS>2.0.CO;2}
}

\end{document}